\documentclass[%
 aip,
 amsmath,amssymb,
 reprint,%
]{revtex4-1}

\usepackage{graphicx,xcolor}
\usepackage{dcolumn}
\usepackage{bm}

\usepackage[utf8]{inputenc}
\usepackage[T1]{fontenc}
\usepackage{mathptmx}
\usepackage{etoolbox}
\usepackage{ulem}

\makeatletter
\def\@email#1#2{%
 \endgroup
 \patchcmd{\titleblock@produce}
  {\frontmatter@RRAPformat}
  {\frontmatter@RRAPformat{\produce@RRAP{*#1\href{mailto:#2}{#2}}}\frontmatter@RRAPformat}
  {}{}
}%
\makeatother
\begin{document}

\preprint{AIP/123-QED}

\title[]{Revisiting the Growth Rate of the Relativistic Tearing Instability: The Role of the Non-ideal MHD Structure}
\author{K. Sugimoto}
\altaffiliation[Present address: ]{SIT Research Laboratories, Shibaura Institute of Technology, Toyosu 3-7-5, Koto-ku, Tokyo 135-8548, Japan}
 \email{kaoru-s@sic.shibaura-it.ac.jp}
\affiliation{Yukawa Institute for Theoretical Physics (YITP), Kyoto University, Kitashirakawa-Oiwakecho, Sakyo-Ku, Kyoto 606-8502, Japan
}%

\author{K. Ioka}
\affiliation{%
Yukawa Institute for Theoretical Physics (YITP), Kyoto University, Kitashirakawa-Oiwakecho, Sakyo-Ku, Kyoto 606-8502, Japan
}%

\author{M. Shibata}
\affiliation{%
Yukawa Institute for Theoretical Physics (YITP), Kyoto University, Kitashirakawa-Oiwakecho, Sakyo-Ku, Kyoto 606-8502, Japan
}%
\affiliation{%
Max Planck Institute for Gravitational Physics (Albert Einstein Institute), Am M\"{u}hlenberg 1, Postdam-Golm 14476, Germany
}%

\date{\today}

\begin{abstract}
Magnetic reconnection in magnetically dominated pair plasmas is a key process in high-energy astrophysical systems. We revisit the relativistic tearing instability in a Harris current sheet and derive an improved analytical expression for its linear growth rate and the most unstable wavenumber. The key modification is the treatment of the vector potential perturbation in the non-ideal magnetohydrodynamic (MHD) region. Instead of the conventional constant-A approximation, we use an extrapolated-A approximation, in which the ideal-MHD solution is linearly extrapolated into the non-ideal region. Comparison with two-dimensional particle-in-cell simulations shows that the revised theory improves the prediction of the most unstable wavenumber. The improvement is most pronounced at low particle  drift velocities, where the particle gyroradius is smaller than the current-sheet thickness and the fastest-growing mode shifts to longer wavelength. The resulting analytical expressions provide an updated benchmark for magnetically dominated reconnection and its applications to high-energy astrophysical plasmas, including gamma-ray bursts and fast radio bursts.
\end{abstract}

\maketitle

%

\section{\label{sec:level1}Introduction:\protect\\ }
Tearing instability is one of the fundamental instabilities that develop in current sheets embedded in magnetized plasmas \cite{Furth63}. Small perturbations in the electric current and reconnecting magnetic field can grow through this instability, leading to changes in magnetic topology and current structure. Because tearing instability occurs over a wide range of plasma parameters, such as density and temperature, in environments ranging from laboratory to space plasmas, it is regarded as a universal phenomenon in magnetized plasmas. 
In the context of fusion plasmas, tearing instability has been extensively studied \cite{Furth73, Seo24} because, in devices such as tokamaks and stellarators, it induces magnetic reconnection accompanied by the formation of magnetic islands, which can significantly degrade plasma confinement. In particular, the subsequent nonlinear stage of tearing mode is closely associated with confinement degradation and major disruptions. Therefore, understanding the underlying physics of tearing instabilities is essential to predict and control plasma behavior in high-performance fusion regimes. 
Tearing instability also plays a crucial role in high-energy astrophysical environments, including tenuous collisionless plasmas in which resistive dissipation is expected to be negligible, because it can trigger magnetic reconnection. 
Through reconnection, magnetic energy is efficiently converted into plasma kinetic, thermal energy, and electromagnetic radiation. 

Over the past several decades, particular attention has been devoted to magnetic reconnection in a magnetically dominated plasma \cite{Zenitani01,Zenitani07,Zenitani08guid,Zenitani08selfreg,Petropoulou19,Guo14}, in which the magnetic energy density exceeds the particle rest-mass energy density. The magnetically dominated reconnection has been proposed as a possible mechanism for powering gamma-ray bursts\cite{McKinney12,Thompson94,Zhang11,Meszaros97} and coherent radio emission \cite{Lyubarsky20,Philippov19,Mahlmann22}. 
In current sheets, tearing instability may compete with other plasma instabilities, such as Kelvin-Helmholtz instability and drift-kink instability. Understanding the growth rate of tearing instability is therefore crucial for evaluating the stability of magnetized plasma systems and for clarifying whether reconnection can drive a wide range of high-energy astrophysical phenomena. 
For this reason, tearing instability and reconnection physics have been investigated extensively for more than half a century, from pioneering studies to the present, within both magnetohydrodynamic (MHD) and kinetic theoretical frameworks \cite{Furth63,Furth73,Hazeltine75,Drake77,Parker57,Petschek64}. In recent years, research have expanded to the unusual regime in which electron-scale kinetic effects play a central role \cite{Phan18,Liu25,Mishra26,Russell26}.

Tearing instability also serves as an important benchmark for numerical simulations of magnetic reconnection. Because well-established theoretical predictions for its linear growth rate exist, comparisons between simulations and theory provide a stringent test of whether a numerical model properly captures the underlying plasma dynamics. For this reason, linear tearing-mode simulations can be used as a standard vertification benchmark for reconnection studies.
In recent years, advances in computational capabilities have made first-principles kinetic simulations of plasma instability increasingly feasible. In particular, modern Particle-In-Cell (PIC) simulations are now capable of investigating reconnection physics in magnetically dominated plasmas with magnetization parameter $\sigma\equiv B^{2}/4\pi Nmc^{2} \gg1$, where $B$, $N$, $m$ and $c$ denote characteristic magnetic-field strength, plasma density, particle rest-mass, and the speed of light, respectively. Such simulations have recently been extended to regimes where radiation damping and pair-production processes play significant roles \cite{Hakobyan19,Chernoglazov23,Zhang24}. This progress has shed light on the possibility of studying the relativistic tearing mode from a fully kinetic perspective where the plasma is magnetically dominated and relativistically hot. Studies aimed at directly determining growth rates through numerical calculations\cite{Demidov24} has also been actively pursued. 

A well-known kinetic derivation of tearing-mode growth rate was originally developed for non-relativistic collisionless current sheets in the Earth’s magnetotail \cite{Coppi66}. In this approach, the Harris equilibrium \cite{Harris62,Hoh66} was adopted as the equilibrium state, and perturbation theory was then applied.
The basic idea is that the magnetic energy stored in the magnetized plasma outside the current sheet is converted into particle kinetic energy through the induced electric fields in the demagnetized central region, and the corresponding rate of energy conversion is interpreted as the instability growth rate. 
This theoretical framework has also been extended to relativistic tearing instabilities \cite{Zelenyi79,Hoshino20}. The resulting theory predicts a maximum growth rate at the most unstable wavenumber $k_{\rm max}$ satisfying $k_{\rm max}w$ = 0.577, where $w$ denotes the current-sheet thickness.
However, discrepancies between the theoretical prediction and PIC simulations have been observed in the most unstable wavenumber. This discrepancy motivates the present re-examination and refinement of the theoretical growth rate expression.

This paper is organized as follows. Section II presents an overview of the conventional growth-rate theories and compares them with the theoretical formula derived in this study. Section III derives the linear growth rate of the tearing instability in magnetically dominated plasma from the energy equations of pertubed electromagnetic fields and currents (see Appendix\,\ref{App_B} for details) and compares the resulting formula with growth rates obtained from PIC simulations. Finally, Section IV provides discussion and conclusions. 

\section{Overview and Comparison between the equations of relativistic tearing instability growth rate}
Throughout this paper, we adopt units such that the Boltzmann constant $k_{B}=1$. A well-known theoretical expression for the growth rate of the tearing instability in magnetically dominated plasmas was derived by Zelenyi and Krasnoselskikh \cite{Zelenyi79}. In their analysis, the plasma is assumed to be in a Harris equilibrium\cite{Harris62}, i.e., a one-dimensional current-sheet equilibrium in which oppositely directed magnetic fields are sustained by the current carried by oppositely drifting charged particles. In such an equilibrium, pressure balance requires the plasma pressure inside the current sheet to balance the magnetic pressure outside it. Since that implies $NT\sim B^{2}/8\pi$, where $T$ is a characteristic plasma temperature, one obtains $T/mc^{2}\sim \sigma$. Consequently, for magnetically dominated plasmas with a magnetization parameter $\sigma\ge 1$, the current-sheet plasma is typically relativistically hot, with thermal energies comparable to or exceeding the particle rest-mass energy.  
In this paper, we use the term "magnetically dominated" when referring to $\sigma>1$, reserving "relativistically hot" and "relativistic-drift" for thermal and drift relativistic effects, respectively. 

To derive the growth-rate formula\cite{Zelenyi79}, Zelenyi and Krasnoselskikh further assume that the associated drift velocity is much smaller than the speed of light $c$. Their theoretical framework is not limited to electron-positron plasmas; rather, it can be applied more generally to electron-ion plasmas with arbitrary mass ratios, temperature ratios, and charge states. The growth rate is therefore derived in a unified framework applicable to a broad range of plasma compositions. As described in the Appendix \ref{App_B}, the growth rate is derived from the balance of energy conversion between the perturbed electromagnetic fields and currents in the inner and outer regions of the current sheet. In the outer magnetized region (ideal MHD region), tearing-mode eigenfunctions \cite{White86} are employed as perturbations, whereas in the inner unmagnetized region (non-ideal MHD region), the perturbed vector potential is assumed to be spatially uniform across the inner layer, corresponding to the constant-A approximation, because the thickness of this region could be sufficiently thin. Under these assumptions, the growth rate $\gamma$ of an unstable mode with wavenumber $k$ in a pair-plasma is given by 
\begin{eqnarray}
\gamma\left(k\right)=\frac{2\sqrt{2}}{\pi}\frac{kw}{\tau_{c}}\left(1-k^{2}w^{2}\right)\beta^{3/2}
\label{eq:one},
\end{eqnarray}
where $\tau_{c}=w/c$ is the characteristic timescale for light to traverse the current-sheet thickness, and $\beta$ denotes the average particle drift speed normalized by $c$.
A limitation of this expression is that the predicted growth rate increases monotonically with increasing drift velocity. At sufficiently large drift speeds, however, relativistic-drift effects are expected to increase the particle inertia and thereby lengthen the particle transit time across the current sheet, delaying current dissipation and suppressing the growth rate.

This issue was addressed by Hoshino \cite{Hoshino20}, who derived a modified theoretical expression that is applicable even when the drift speed approaches the speed of light. His analysis focused on high-energy pair plasma environments consisting of electrons and positrons. The revised growth-rate formula is given by
\begin{eqnarray}
\gamma\left(k\right)=\frac{2\sqrt{2}}{\pi}\frac{kw}{\tau_{c}}\left(1-k^{2}w^{2}\right)\frac{\beta^{3/2}}{\Gamma_{\beta}},
\label{eq_growth_Hoshino}
\end{eqnarray}
where $\Gamma_{\beta}=1/\sqrt{1-\beta^{2}}$ denotes the Lorentz factor associated with the plasma bulk drift.
In that study, theoretical predictions were compared with results obtained from two-dimensional PIC simulations, and the simulated maximum growth rate was found to be in good agreement with the theoretical values. Although this theoretical expression was originally derived under the assumption that the particle gyroradius is sufficiently smaller than the current-sheet thickness, recent studies have shown that it remains valid even when the ratio approaches unity \cite{Shoeffler19}.

However, it has also been observed that the simulated most unstable wavenumber does not agree with the theoretical prediction. According to Eq.\,\eqref{eq_growth_Hoshino}, the maximum growth rate always attained at the wavenumber satisfying $kw=1/\sqrt{3}\simeq 0.577$, consistent with the earlier result obtained by Zelenyi and Krasnoselskikh. In contrast, PIC simulations, spanning drift velocities of $0.05\le\beta\le0.95$, indicate that the growth rate peaks at $kw \simeq 0.39$ or lower\cite{Hoshino20,Shoeffler25}. Thus, the theoretical prediction and simulation results differ by approximately $30\%$ in the most unstable wavenumber.

These discrepancies may be resolved by reevaluating the energy equation after correcting the spatial structure of the perturbation in the non-ideal MHD region. In fact, in subsequent work by Hoshino\cite{Hoshino21}, theoretical expressions were derived without invoking the constant-A approximation, and the resulting predictions agreed with two-dimensional PIC simulations not only in the magnitude of the growth rate but also in the most unstable wavenumber. 
The essence of this modification is that, instead of assuming a spatially constant perturbation in the non-ideal MHD region, the perturbative solution obtained in the ideal MHD region is extrapolated into the non-ideal MHD region when evaluating the energy equation, thereby altering contribution from the inner region to the growth rate (see the Appendix \ref{App_B} for more details).
In this paper, we use the term extrapolated-A approximation to refer to this approach.
Since the aforementioned study\cite{Hoshino21} considered only plasmas with non-relativistic temperatures, it remains important to investigate whether this correction method is also applicable to relativistically hot plasmas when evaluating the growth rate of tearing instabilities.

Using the extrapolated-A approximation, we derive a modified growth rate expression for the tearing instability of magnetically dominated plasma in this paper. The resulting growth rate is written as
\begin{eqnarray}
\gamma\left(k\right)=\frac{2\sqrt{2}}{\pi}\frac{kw}{\tau_{c}}\frac{1-k^{2}w^{2}}{kw+\sqrt{r_{g}/w}}\frac{\beta^{3/2}}{\Gamma_{\beta}}
\label{eq_growth_New}
\end{eqnarray}
where $r_{g}$ denotes the gyroradius of a thermal particle evaluated using the asymptotic magnetic field strength in the Harris equilibrium. The present expression is applicable to relativistically hot plasmas and remains valid for drift velocities spanning the non-relativistic to relativistic regimes. According to this modification, the most unstable wavenumber becomes small. Assuming that the plasma thermal energy is much larger than the particle rest-mass energy, we approximate the thermal Lorentz factor as $\gamma_{\rm th}=T/mc^{2}$. The gyroradius is then written as $r_{g}\sim\gamma_{\rm th}mc^{2}/eB$, where $e$ denotes the elementary charge. Using the Harris equilibrium condition $2T/w=e\beta B$, the ratio of the gyroradius to the Harris-sheet thickness becomes $r_{g}/w\sim\beta/2$.
As a consequence of this modification, the most unstable wavenumber becomes explicitly dependent on the average particle drift speed $\beta$.

\section{Modification of the Theoretical Growth Rate and Comparison with PIC Simulations}
In this section, we describe the modification of the theoretical growth rate of the tearing instability in magnetically dominated plasma and present the setup and results of the PIC simulations used for comparison. As the equilibrium configuration, we consider a Harris current sheet (normal to the $x$-direction) with antiparallel magnetic (in the $y$-direction) and an electric current flowing in the $z$-direction. We then derive the growth rate of the tearing instability arising from this equilibrium. The main steps of the derivation are summarized below, while the detailed calculations are deferred to Appendix \ref{App_B}. All physical quantities are assumed to be uniform along the z-direction, i.e., $\partial/\partial z=0$.
The plasma is assumed to consist of electrons and positrons. 

Following previous studies, we begin with the energy equation describing the interaction between electromagnetic fields and plasmas:
\begin{eqnarray}
\frac{\partial}{\partial t}\left(\frac{B^{2}}{8\pi}\right)+\nabla\cdot\left(\frac{c}{4\pi}\boldsymbol{E}\times\boldsymbol{B}\right)=-\boldsymbol{E}\cdot\boldsymbol{J}
\label{eq_elemag_energy}.
\end{eqnarray}
To linearize the equation, we introduce a perturbation of the form $\phi_{1}=\tilde{\phi}(x)\exp\left(iky-i\omega t\right)$ for an arbitrary physical quantity $\phi$, where $\tilde{\phi}\left(x\right)$ is the mode-amplitude associated with the Fourier mode ${\rm exp}\left(iky-i\omega t\right)$. We then perform spatial integrations over $x=-\infty\sim\infty$ and $y=-\pi/k\sim\pi/k$ on both sides of the equation. Since the perturbations should vanish as $|x|\rightarrow\infty$, the volume integral of the second term on the left-hand side can be transformed into a surface integral at infinity, which vanishes. The first-order perturbation terms vanish after integration over the periodic domain in the $y$-direction, leaving only second-order terms. Separating the adiabatic and non-adiabatic contributions, we obtain
\begin{equation}
    \begin{aligned}
          \int\limits_{y=-\frac{\pi}{k}}^{y=\frac{\pi}{k}}\int\limits_{x=-\infty}^{x=\infty}
          \left\{\frac{\partial}{\partial t}\left(\frac{\boldsymbol{B}_{1}\cdot\boldsymbol{B}_{1}}{8\pi}\right)+\left(\boldsymbol{E}_{1}\cdot\boldsymbol{J}_{1}\right)_{\rm ad}\right\}dxdy\\
         =-\int\limits_{y=-\frac{\pi}{k}}^{y=\frac{\pi}{k}}\int\limits_{x=-r_{m}}^{x=r_{m}}\left(\boldsymbol{E}_{1}\cdot\boldsymbol{J}_{1}\right)_{\rm non-ad}dxdy.
    \end{aligned}
    \label{energy_eq_divided}
\end{equation}
We assumed that non-adiabatic energy conversion occurs within the meandering region centerd on the current sheet. Here, $r_{m}$ denotes the meandering radius, which is given by $\sqrt{r_{g}w}$ (see Appendix \ref{App_A} for details).
The perturbation of the electromagnatic field can be expressed in terms of the perturbation of the vector potential. The spatial distribution of the perturbation $\tilde{A}_{z1}(x)$ in the ideal-MHD region can be written as follows \cite{White86},
\begin{equation}
    \tilde{A}_{z1}(x)=\tilde{a}_{k}\left(1+\frac{\tanh\left(|x|/w\right)}{kw}\right)\exp\left(-k|x|\right) 
    \label{ideal_mhd_A},
\end{equation}
where $\tilde{a}_{k}$ denotes the amplitude of perturbed vector potential. Previous studies \cite{Zelenyi79, Hoshino20} assumed that the perturbation in the non-ideal MHD region is uniform profile in the $x$-direction. Considering the connection between the vector potential in the ideal MHD region and that in the non-ideal MHD region, the slope of $\tilde{A}_{z1}(x)$ near $x=0$ becomes shallow as $kw\rightarrow1$, in which case the constant-A approximation is valid. In contrast, as $kw\rightarrow0$, the slope of $\tilde{A}_{z1}(x)$ near $x=0$ becomes steep, which enhances the slope-discontiuity relative to the constant solution in the non-ideal MHD region.
To resolve this discontinuity, we extrapolate the solution obtained in the ideal MHD region into the non-ideal MHD region, following previous analyses of tearing instability in plasmas with non-relativistic temperatures\cite{Hoshino21}.
At the boundary between the ideal MHD and the non-ideal MHD regions, we perform a Taylor expansion of the outer solution and approximate it by the linar relation $\tilde{A}_{z1}(x)\propto 1+x/kw^{2}$, which is then extrapolated into the non-ideal MHD region \cite{Hoshino21}. Here, we assume that the meandering radius $r_{m}$ is much smaller than the current-sheet thickness $w$. Under this assumption, tanh$\left(x/w\right)$ and exp$\left(-kx\right)$ are expanded in Taylor series, and only the lowest-order terms are retained. By integrating Eq.\,(\ref{energy_eq_divided}), we obtained the modified growth-rate formula given in Eq.\,(\ref{eq_growth_New}). The detailed derivation is provided in Appendix \ref{App_B}.

To verify the newly derived growth rate, we performed two-dimensional PIC simulations in the $x$-$y$ plane using the PICLS code \cite{Sentoku08}. As the initial condition, we adopted a relativistic Harris equilibrium \cite{Hoh66} composed of an electron–positron pair plasma. 
The spatial profiles of the equilibrium quantities are given by the magnetic field
\begin{eqnarray}
\boldsymbol{B}(x)=B_{0}\tanh\left(x/w\right)\mathbf{e}_{y},
\end{eqnarray}
and the particle number density of each species (electrons and positrons) is given by
\begin{eqnarray}
n_{s}(x)=n_{0}\cosh^{-2}\left(x/w\right)+n_{b},
\end{eqnarray}
where the subscript $s(=e,p)$ denotes the particle species.
Here, $n_{b}$ represents a uniform, non-drifting background particle density and is set to $n_{b}=0.05n_{0}$. In the present simulations, the plasma is assumed to be isothermal for both particle species. The temperatures of the drifting component and background components in the laboratory frame are fixed at $T_{e,p}=10m_{e}c^{2}$, where $m_{e}$ is the electron rest-mass, throughout all runs. 
For simplicity, the mean drift velocities of electrons and positrons are assumed to have equal magnitudes and opposite directions, thereby ensuring initial charge neutrality of the Harris profile. Positrons drift in the negative $z$-direction.
Under these conditions, simulations were performed for three cases: $\beta=$0.3, 0.5 and 0.8.

\begin{figure}
\includegraphics[width=8.5cm]{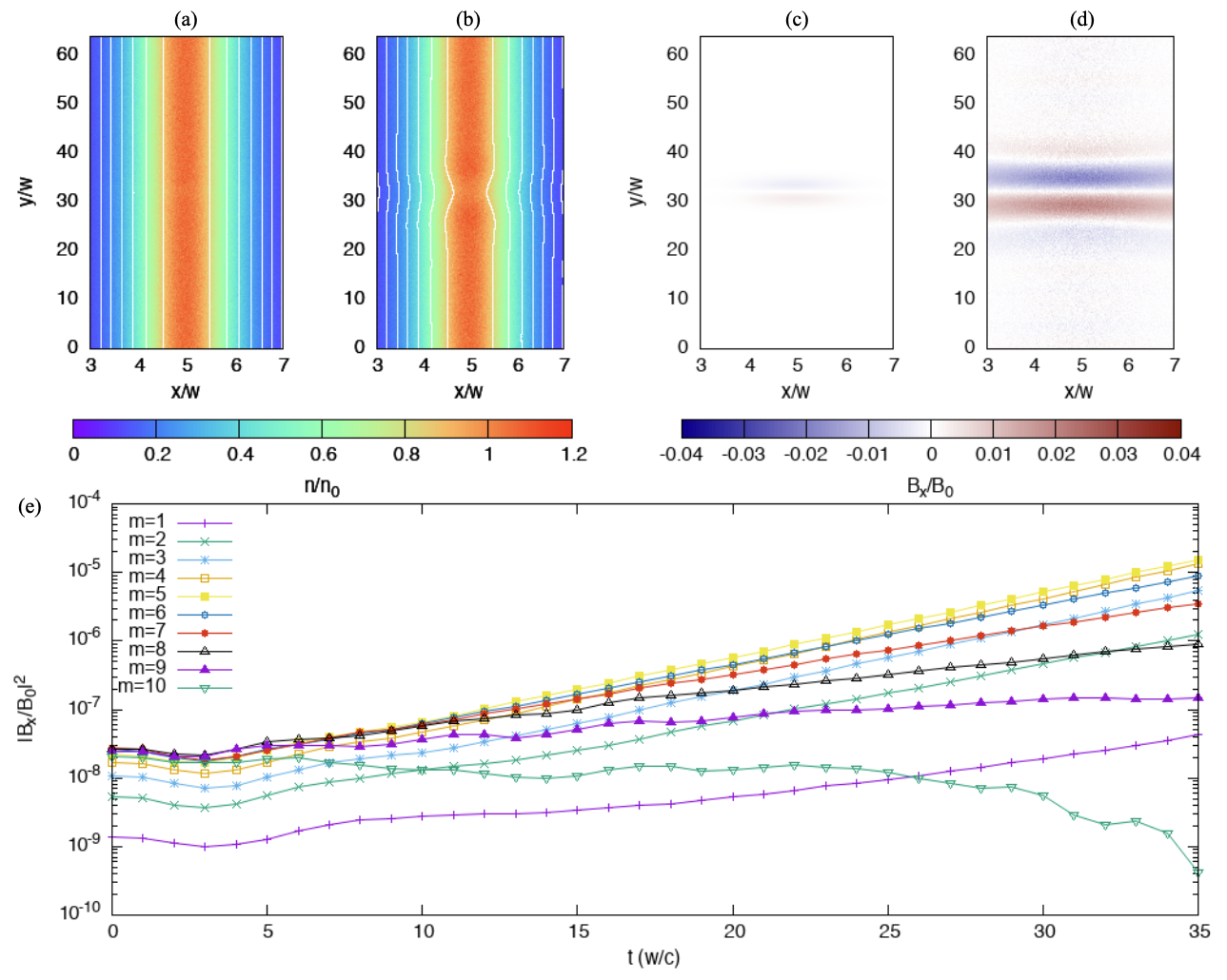}
\caption{\label{fig:epsart} Evolution of the tearing-mode instability obtained from the two-dimensional PIC with electron and positron temeratures $T_{e,p}=10m_{e}c^{2}$ and $\beta=0.8$. $\beta$ represents the average particle drift speed normalized by $c$. Panels (a) and (b) show the positron number density at $t=0$ and $t=30w/c$, respectively. The white solid curves represent magnetic field lines. Panels (c) and (d) show the corresponding spatial distributions of the magnetic-field component $B_x$. The initial perturbation imposed at the center of the current sheet grows with time, accompanied by a decrease in particle density and distortion of the magnetic-field structure around the dissipation region. Panel (e) shows the temporal evolution of the squared Fourier amplitudes, $|\tilde{B}_x(k_{\rm m})/B_{0}|^2$, for modes $m=1-10$, where $k_m=2\pi {\rm m}/L_y$ and ${\rm m}$ indicates an integer. The growth rates are determined from the linear growth phase of each mode.}
\label{evo}
\end{figure}

In the PICLS code, the maximum positron (electron) density is normalized to $n_{0}=1$. 
The spatial and temporal grid spacings satisfy $\Delta x=\Delta y=c\Delta t$. 
In all simulations, the numerical domain has dimensions of $L_{x}=10w$ in the $x$-direction and $L_{y}=64w$ in the $y$-direction. For the cases of $\beta=0.5$ and 0.8, the current-sheet thickness was set to $w=200\Delta x$. 
For $\beta=0.3$, the growth rate is relatively small and therefore more susceptible to numerical noise; accordingly, a finer resolution of $w=800\Delta x$ was adopted. 
We confirmf that the simulation results are converged with respect to the spatial and temporal resolutions used in the simulations.
Periodic boundary conditions are applied in the $y$-direction, while conducting-wall boundary conditions are imposed at the boundaries in the $x$-direction. To trigger tearing instability, a small initial magnetic perturbation is imposed. 
In order to localize the perturbation near the center of the current sheet, we assume the following initial perturbed vector potential: 
\begin{equation}
    A_{z1}\propto \exp\left(-\frac{x^{2}}{l_{x}^{2}}-\frac{y^{2}}{l_{y}^{2}}\right)\cosh^{2}\left(\frac{x}{l_{x}}\right).
\end{equation}
The corresponding perturbed magnetic field and perturbed current are introduced in the PIC simulations, and the initial particle velocities are adjusted such that the perturbed current satisfies the steady-state Amp\`ere's law, $\nabla\times\boldsymbol{B}_{1}=4\pi\boldsymbol{J}_{1}/c$. 
Here, the characteristic scales are chosen as $l_{x}=w$ and $l_{y}=2w$, respectively. The amplitude of the initial perturbed magnetic field $B_{x1}$ is set such that its maximum value is $0.01B_{0}$.

As a representative case, Fig.\ref{evo} presents the evolution of the tearing-mode instability obtained from a two-dimensional PIC simulation with electron and positron temperatures $T_{e,p}=10m_{e}c^{2}$ and $\beta=0.8$. Fig.\ref{evo}(a) and \ref{evo}(b) show the positron number density at $t=0$ and $t=30w/c$, respectively. Although the simulation domain extends from $x=0$ to $x=10w$, only the region around the current sheet is enlarged for clarity. The white solid curves indicate the magnetic field lines at the corresponding times.
As time progresses, the particle density decreases around the center of the current sheet ($x=5w,; y=32w$), indicating dissipation of the current. Accompanying this structural change, the magnetic field lines surrounding the dissipation region become increasingly distorted.

Figure \ref{evo}(c) and \ref{evo}(d) show the spatial distribution of the $x$-component of the magnetic field, $B_x$, at $t=0$ and $t=30w/c$, respectively. The localized structure seen at the center of the current sheet ($x=5w,; y=32w$) at $t=0$ corresponds to the initial magnetic perturbation introduced in the simulation. Similar to the density evolution, the perturbation grows around the center of the current sheet, exhibiting both an increase in amplitude and an expansion of the perturbed region.
In addition, a positive $B_x$ component appears in the region $y>38w$, while a negative component appears in the region $y<26w$. These components have signs opposite to that of the original perturbation field. From the ratio between the spatial period over which the sign reversal occurs and the size of the simulation box, we infer that a mode with a wavenumber of approximately $k\simeq(2\pi/L_y)\times 5$ has become dominant.

The procedure used to extract the growth rate from the PIC simulation results is as follows. 
First, we compute the average $\overline{B}_{x}(y)$ of the $x$-component of magnetic field, $B_{x}(x,y)$, over the range $|x|\le0.1w$. 
We then calculate the Fourier coefficient corresponding to the Fourier mode ${\rm exp}\left(ik_ my\right)$, where $k_m=2\pi m/L_{y}$ and $m$ indicates an integer. Figure \ref{evo}(e) shows the temporal evolution of the squared magnitude of the Fourier coefficients, $|\tilde{B}_x(k_m)/B_{0}|^2$, for modes with $m=1$--$10$ extracted using the above procedure. For modes with $m=1$--$9$, a phase of approximately exponential growth is clearly observed, appearing as a linear increase in the semi-logarithmic representation. The growth rate of each mode was determined from the slope of this linear phase using the least-squares method.
In contrast, the $m=10$ mode does not exhibit growth and instead decays with time. This behavior may be attributed to two factors. First, the growth rate of the $m=10$ mode is relatively small compared with those of the other modes, making it more susceptible to numerical noise. Second, because the wavelength of this mode is short, the structure is more difficult to resolve accurately with a finite number of grid cells and is therefore more easily disrupted by numerical effects. For modes for which no clear growth could be identified, the growth rate was taken to be zero.

\begin{figure}
\includegraphics[width=8.0cm]{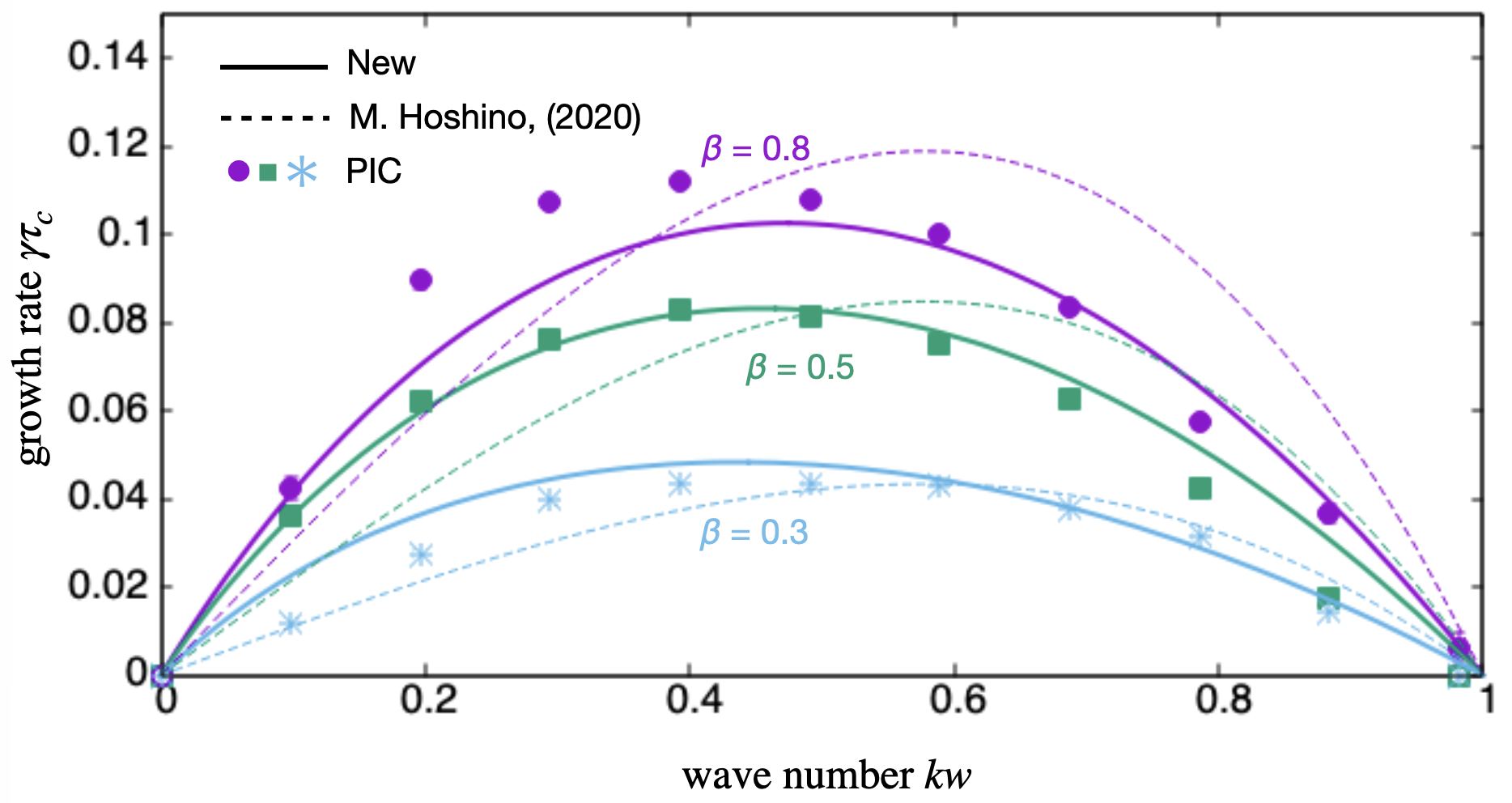}
\caption{\label{fig:epsart} Wavenumber dependence of the growth rate of the relativistic tearing instability measured from PIC simulations with electron and positron temperatures $T_{e,p}=10m_{e}c^{2}$. Here, $w$ is the current-sheet thickness, $\beta$ is the bulk drift speed normalized by the speed of light, and $\tau_{c}$ is the light-crossing time across the current sheet. The symbols denote results obtained from the simulations, with different marker shapes corresponding to different bulk drift speed. The solid and dashed curves represent Eq.\,\eqref{eq_growth_New} and Eq.\,\eqref{eq_growth_Hoshino}, respectively.}
\label{comp_growth}
\end{figure}

Figure \ref{comp_growth} compares the simulation results with the theoretical predictions. 
Two theoretical curves are shown: the growth rate obtained in previous studies (dashed line; Eq.\,\eqref{eq_growth_Hoshino}), and that derived in the present work (solid line; Eq.\,\eqref{eq_growth_New}). 
Compared with conventional theoretical expressions, the present formulation reproduces the most unstable wavenumber more accurately, indicating that the spatial structure of perturbations in the inner non-ideal MHD region plays a crucial role. In principle, the inner solution should asymptotically connect to the outer ideal MHD solution with linear behavior \cite{White86}, and the approximate solution obtained here satisfies this requirement.
In comparing these results, it should be noted that the theoretical values are multiplied by an empirical factor of 0.8, whereas a factor of approximately 0.7 was applied in the previous study \cite{Hoshino20}. This discrepancy may be attributable to finite-Larmor-radius effects not included in the theory. Although several possible origins of this discrepancy have been proposed, no consensus has yet been reached. 
According to the newly derived growth rate, the wavenumber corresponding to the maximum growth depends on the drift velocity $\boldsymbol{\beta}$, with smaller values of $\beta$ yielding a smaller most unstable wavenumber. A smaller $\beta$ implies that the gyroradius is small compared with the current-sheet thickness. In this limit, kinetic scale effects become less pronounced and the system approaches a fluid-like reconnection regime, in which a fluid descreption such as the Sweet-Parker or Petschek models become applicable. At the same time, the discrepancy between Eqs.\,\eqref{eq_growth_Hoshino} and \eqref{eq_growth_New} in predicting the wavenumber of the most-unstable mode becomes more pronounced, with the difference increasing as the system approaches the fluid limit.

\section{Conclusions}
Tearing instability is one of the fundamental mechanisms driving spontaneous magnetic reconnection in both fusion and astrophysical plasmas. In particular, the magnetic reconnection in magnetically dominated plasma, characterized by a magnetization parameter much greater than unity, has been proposed as a possible trigger of explosive high-energy phenomena in astrophysical systems\cite{Wada20,Carrasco20,Carrasco21}. Meanwhile, laboratory experiments using high-intensity lasers are expected to advance the understanding of the relativistic magnetic reconnection \cite{Huang23}.

In this study, we revisited the wavenumber dependence of the growth rate of the relativistic tearing instability by comparing analytical predictions with PIC simulations. Previous studies derived the growth rate under the assumption that the perturbed vector potential is spatially uniform within the non-ideal MHD region. However, discrepancies have been reported between the theoretically predicted most unstable wavenumber and results obtained from first-principles PIC simulations. To address this issue, we approximated the perturbed vector potential in the non-ideal MHD region by linearly extending the solution from the ideal MHD region and reevaluated the growth rate accordingly. Unlike the conventional constant-$A$ approximation, which removes the wavenumber dependence of the perturbed vector potential in the non-ideal region, the extrapolated-A approximation retains this dependence.
To validate the revised theoretical expression, we performed two-dimensional PIC simulations under magnetically dominated plasma conditions. 
The comparison shows that the revised model predicts the most unstable wavenumber more accurately than conventional formulae. The present results provide a quantitative benchmark for testing and validating numerical codes aimed at relativistic tearing instability simulations.

As the drift velocity in the current sheet decreases, the most unstable wavenumber becomes smaller. This corresponds to a regime in which the current sheet thickness becomes large compared with the particle gyroradius, approaching conditions in which the instability structure becomes closer to that predicted by fluid models.
These findings indicate that an accurate description of the wavenumber dependence of the growth rate requires careful treatment of the spatial structure of perturbations within the non-ideal MHD region.
The present analysis is limited to a relatively fundamental regime of tearing instability. In realistic relativistic environments, additional effects, such as radiative cooling of charged particles and density modification due to pair production, are expected to play important roles. Although previous studies have examined linear growth rates including radiative damping, the explored parameter range remains limited, leaving considerable room for further investigation. Moreover, once the instability enters the nonlinear stage, important processes include magnetic-energy release through magnetic reconnection and plasmoid dynamics. In particular, precursor emission associated with binary neutron star mergers\cite{Wada20,Carrasco20,Carrasco21} and driven by plasmoid coalescence has been discussed in previous studies\cite{Philippov19,Most20}; however, its validity in relativistic regimes with strong radiative and pair-production effects remains unclear. These issues remain important subjects for future work.

\begin{acknowledgments}
The authors would like to thank M. Iwamoto, S. F. Kamijima, Y. Takei, R. Kuze, R. Nishiura, W. Ishizaki, M. Hata, T. Sano, and M. Hoshino for fruitful discussions. We are also grateful to Y. Sentoku for providing the computational code used in this study, as well as for valuable discussions and suggestions regarding the simulation setup. The simulations were carried out using the supercomputer Yukawa-21 provided by Yukawa Institute for Theoretical Physics at Kyoto University, and the supercomputer SQUID provided by D3 center at the University of Osaka.
This work was supported by JSPS KAKENHI Grant Number 23H04900, 22H00130, 23H05430, 23H01172, 26H02045, 23K20038. The authors are grateful to the YITP long-term workshop ``Multi-Messenger Astrophysics in the Dynamic Universe'' held at the Yukawa Institute for Theoretical Physics, Kyoto University, for stimulating discussions that contributed to this work.

\end{acknowledgments}

\section*{Conflict of Interest}

The authors have no conflicts to disclose.

\section*{Author Contributions}

K.S.: Conceptualization (lead); Investigation (lead); Formal analysis (lead); Writing -- original draft (lead).

K.I.: Conceptualization (supporting); Writing -- review \& editing (supporting); Funding acquisition (equal).

M.S.: Conceptualization (supporting); Writing -- review \& editing (supporting); Funding acquisition (equal).

\section*{Data Availability Statement}
The data that support the findings of this study are available from the corresponding author upon reasonable request.

\appendix

\section{Meandering radius of relativistic particles}\label{App_A}

We consider the motion of a charged particle (rest-mass $m_{0}$, charge $q>0$) in an antiparallel magnetic field configuration described by the Harris equilibrium, and evaluate the width of the non-ideal MHD region, defined as the spatial extent of unmagnetized particle orbits in the $x$-direction. 
For this system, the relativistic energy $E=\gamma_{p} m_{0}c^{2}$, where $\gamma_{p}$ is the particle's Lorentz factor, the canonical momentum in the $y$-direction $P_y=p_y$, and the canonical momentum in the $z$-direction $P_z=p_z+qA_{z}(x)/c$ are conserved. Introducing the four-velocity $u$, we note that $u_y$ is therefore constant. Conserving of the relativistic energy further implies that $u_x^2+u_z^2$ is also constant. Expressing the canonical momentum in the $z$-direction in terms of the four-velocity, we introduce the following two constants of motion:
\begin{equation}
    u_{n}\equiv u_{z}+\frac{q}{m_{0}c}A_{z}(x)={\rm const.},
    \label{equation_u_n}
\end{equation}
\begin{equation}
    u_{\perp}^{2}\equiv u_{x}^{2}+u_{z}^{2}={\rm const.}.
    \label{equation_u_perp}
\end{equation}
Next, for analytical simplicity, we approximate the equilibrium vector potential and the associated antiparallel magnetic field as
\begin{equation} 
A_{z}(x)= 
\begin{cases} 
  B_{0}x+\frac{b}{2}{x^{*}}^{2} &(x<-x^{*})\\ 
  -\frac{b}{2}x^{2} &(|x|\le x^{*})\\ 
  -B_{0}x+\frac{b}{2}{x^{*}}^{2} &(x>x^{*}) 
 \end{cases} ,
 \label{approxi_vec}
\end{equation}
\begin{equation}
    B_{y}(x)= 
    \begin{cases} 
    -B_{0} &(x<-x^{*})\\ 
    bx &(b\equiv B_{0}/x^{*}, |x|\le x^{*})\\ 
    B_{0} &(x>x^{*}) \end{cases} .
\end{equation}
where $B_0$ and $x^{*}$are positive constants representing the characteristic magnetic-field strength and the characteristic spatial length of the magnetic field structure, respectively.
Unmagnetized particle orbits within the non-ideal MHD region can be classified into two types: (i) meandering orbits, which cross the midplane ($x=0$), and (ii) non-crossing orbits, which do not. 
The width of the non-ideal region, i.e., the spatial extent of particle motion in the $x$-direction, can be determined by the turning points satisfying $u_{x}=0$.
As the turning points, the condition
\begin{equation}
  u_{z}^{2}=u_{\perp}^{2}
  \label{eq_uz2_u_perp_2}
\end{equation}
is satisfied, yielding
\begin{equation}
  u_{z}=\pm u_{\perp}.
  \label{eq_uz_pm_u_perp}
\end{equation}
Substituting this into the definition of $u_{n}$, we obtain
\begin{equation}
    -\frac{q}{m_{0}c}A_{z}(x)=\pm u_{\perp} - u_{n}
    \label{eq_turn_point}
\end{equation}
We now focus on meandering particles. Since $A_z(0)=0$, we have $u_n=u_z$ at $x=0$. 
Therefore, $|u_n|$ corresponds to the magnitude of the z-component of the four-velocity at the midplane. Furthermore, because $u_x\ne0$ at $x=0$, the inequality $u_{\perp}>|u_n|$ follows. Since both $u_{\perp}$ and $u_n$ are constants of motion, this relation holds throughout the particle trajectory.
Let $x_t$ denote the $x$-coordinate of the turning point for a meandering particle. 
We assume that the turning points satisfying $u_x=0$ lie within the region $|x|\le x^{*}$. Solving Eq. \eqref{eq_turn_point} for $x_{t}$ using Eq.\eqref{approxi_vec}, we obtain
\begin{equation}
    x_{t}=\pm 2d\sqrt{\frac{m_{0}cu_{\perp}}{qb}},
    \label{turn_point}
\end{equation}
where
\begin{equation}
    d\equiv\sqrt{\frac{1-u_{n}/u_{\perp}}{2}}
\end{equation}
satisfying $0<d<1$. 
The negative branch in Eq. \eqref{eq_turn_point} is not physically admissible. Since $A_{z}(x)\le 0$, the left-hand side satisfies $-qA_z/m_{0}c\ge0$, and therefore the right-hand side must also be non-negative. Under the condition $u_{\perp} > |u_{n}|$, this requirement is satisfied only when the branch $u_{\perp}-u_{n}$ is chosen.
The expression under the square root in Eq. \eqref{turn_point} can be rewritten as
\begin{equation}
    \frac{m_{0}cu_{\perp}}{qb}=\frac{\gamma_{p} m_{0}cv_{\perp}}{qB_{0}}x^{*}{\equiv}r_{g}x^{*}
\end{equation}
where $r_g$ is the relativistic gyroradius. Interpreting $x^{*}$ as the characteristic current-sheet thickness $w$ in the Harris equilibrium, the turning distance (meandering width) scales as
\begin{equation}
    |x_{t}|\sim\sqrt{r_{g}w}
    \label{eq_rela_meandering}
\end{equation}
In this study, this meandering width is used as the characteristic width of the non-ideal region in deriving the instability growth-rate.

\section{Derivation of the growth rate}\label{App_B}
The current sheet is assumed to be in a Harris equilibrium characterized by a relativistic shifted-Maxwellian distribution function \cite{Hoh66},
\begin{equation}
\begin{aligned}
      f_{s0}=&\frac{\overline{n}_{s0}}{4\pi m_{s}^{2}c\Theta_{s} K_{2}\left(m_{s}c^{2}/\Theta_{s}\right) }\times\\
      &\exp\left\{ -\frac{E_{s}-c\beta_{s}p_{z}-q_{s}\beta_{s}A_{z0}\left(x\right) }{ T_{s} } \right\}
\end{aligned}
\label{eq_rela_distribution}.
\end{equation}
where the subscript $s$ denotes the particle species, and the subscript 0 indicates equilibrium quantities. Here, $\overline{n}_{s0}$, $\Theta_s$, and $T_s$ represent the peak particle number density in the bulk rest frame, the temperature measured in the comoving frame, and the temperature measured in the laboratory frame, respectively. $\Theta_{s}$ and $T_{s}$ satisfy the relation $T_{s}=\Theta_{s}/\Gamma_{\beta}$\cite{Kirk03}. $K_2$ denotes the modified Bessel function of the second kind.
Furthermore, $E_s=\sqrt{m_{s}^{2}c^{4}+p^{2}c^{2}}$ is the relativistic particle energy, where $p$ denotes the magnitude of the particle momentum. $q_s$ is the particle charge, $c\beta_s$ is the mean drift velocity of species, $p_z$ is the z-component of the four-momentum, and $A_{z0}$ is the equilibrium vector potential corresponding to the Harris-type antiparallel magnetic field configuration.
We assume that an arbitrary perturbed quantity $\phi$ takes the form
\begin{eqnarray}
\phi_{1}(x,y,t)=\tilde{\phi}_{1}(x)\exp\left(iky-i\omega t\right),
\label{eq_perturbation}
\end{eqnarray}
where $\tilde{\phi}_{1}(x)$ is, in general, a complex-valued function. 
In the present analysis, we assume that the z-component of the perturbed vector potential is dominant. The perturbation is therefore taken to be of the form $\boldsymbol{A}_{1}=A_{z1}\mathbf{e}_{z}=\tilde{A}_{z1}(x)\exp\left(iky-i\omega t\right)\mathbf{e}_{z}$. Here, $\omega={\rm Re}\left[\omega\right]+i{\rm Im}\left[\omega\right]$, and the imaginary part Im\,[$\omega$] corresponds to the growth rate $\gamma$.

Integrating Eq.\,\eqref{eq_elemag_energy} over one wavelength in the $y$-direction and over the entire domain in the $x$-direction yields
\begin{equation}
\begin{aligned}
         \frac{\partial}{\partial t}\int\limits_{y=-\frac{\pi}{k}}^{y=\frac{\pi}{k}}\int\limits_{x=-\infty}^{x=\infty}&\left(\frac{\boldsymbol{B}_{1}\cdot\boldsymbol{B}_{1}}{8\pi}\right)dxdy\\
         &=-\int\limits_{y=-\frac{\pi}{k}}^{y=\frac{\pi}{k}}\int\limits_{x=-\infty}^{x=\infty}\boldsymbol{E}_{1}\cdot\boldsymbol{J}_{1}dxdy.  
\end{aligned}
\label{eq_energy_conservation}
\end{equation}
In deriving this expression, the terms associated with the Poynting vector are transformed into surface integrals evaluated at infinity. 
These terms vanish because the perturbations are assumed to decay sufficiently rapidly at large distances.
The integrand on the right-hand side can be rewritten as 
\begin{equation}
    \begin{split}
        \boldsymbol{E}_{1}\cdot\boldsymbol{J}_{1}&=\frac{\exp\left(2\gamma t\right)}{4}\times\\
        &\left\{\tilde{E}_{z1}\left(x\right)\exp\left(iky-i{\rm Re}[\omega]t\right)+\tilde{E}_{z1}^{*}\left(x\right)\exp\left(-iky+i{\rm Re}[\omega]t\right)\right\}\\
        &\left\{\tilde{J}_{z1}\left(x\right)\exp\left(iky-i{\rm Re}[\omega]t\right)+\tilde{J}_{z1}^{*}\left(x\right)\exp\left(-iky+i{\rm Re}[\omega]t\right)
        \right\}\\
      &=\frac{\exp\left(2\gamma t\right)}{2}{\rm Re}\left[\tilde{E}_{z1}\tilde{J}_{z1}\left(x\right)\exp\left(-i2ky+i2{\rm Re}[\omega]t\right)\right.\\
      &\ \ \ \left. +\tilde{E}_{z1}^{*}(x)\tilde{J}_{z1}(x)\right]
    \end{split}
\end{equation}
Here, the superscript $^*$ denotes the complex conjugate. Likewise, the integrand on the left-hand side can be rewritten as
\begin{equation}
    \begin{split}
        \boldsymbol{B}_{1}\cdot\boldsymbol{B}_{1}&=\frac{\exp(2\gamma t)}{2}{\rm Re}\left[\right.\tilde{\boldsymbol{B}}_{1}(x)\cdot\tilde{\boldsymbol{B}}_{1}(x)\exp\left\{-i2ky+i2{\rm Re}[\omega]t\right\}\\
        & +\tilde{\boldsymbol{B}}_{1}^{*}(x)\cdot\tilde{\boldsymbol{B}}_{1}(x)\left.\right]
    \end{split}
\end{equation}
Terms proportional to $\exp(iky-i{\rm Re}[\omega] t)$ (or $\exp(-iky+i{\rm Re}[\omega] t)$) vanish upon integration over $y$ over one period, $-\pi/k < y < \pi/k$, and are therefore omitted in the following discussion. Furthermore, the remaining integration over $y$ merely yields a common factor of $2\pi/k$ on both sides, which can be canceled out. Consequently, Eq.~\eqref{eq_energy_conservation} can be written as
\begin{equation}
    \begin{split}
            \frac{\partial}{\partial t}\int\limits_{x=-\infty}\limits^{x=\infty}\frac{|\tilde{\boldsymbol{B}}_{1}(x)|^{2}}{8\pi}&\exp\left(2\gamma t\right)dx=\\
            &-\int\limits_{x=-\infty}\limits^{x=\infty}{\rm Re}\left[\tilde{E}_{z1}^{*}(x)\tilde{J}_{z1}(x)\right]dx
    \end{split}
\end{equation}
The resulting expression can be decomposed into ideal- and non-ideal MHD contributions as
\begin{equation}
   \begin{split}
   \int\limits_{x=-\infty}^{x=\infty}\left(\frac{\partial}{\partial t}\left(\frac{|\tilde{\boldsymbol{B}}_{1}(x)|^{2}}{8\pi}{\rm e}^{2\gamma t}\right)+{\rm Re}\left[\tilde{E}_{z1}^{*}(x)\tilde{J}_{z1}^{\rm ideal}(x)\right]{\rm e}^{2\gamma t}\right)dx\\
   =-\int\limits_{x=-r_{m}}^{x=r_{m}} {\rm Re}\left[\tilde{E}_{z1}^{*}(x)\tilde{J}_{z1}^{\rm non-ideal}(x)\right]{\rm e}^{2\gamma t}dx
    \end{split}
    \label{energy_eq}
\end{equation}
For the non-ideal MHD contribution, we assume that particle acceleration by the electric field acting on unmagnetized particles is the dominant process, and therefore neglect the contribution from the perturbed magnetic energy density. 
In addition, we consider the regime where the electromagnetic perturbations satisfy ${\rm Re}[\omega]\sim0$, ${\rm Im}[\omega] \ll v_{\mathrm{th}}k$ and $k\lambda_D < 1$. Under these conditions, the displacement-current term in Amp{\`e}re's law can be neglected. Furthermore, the instability growth rate remains much smaller than the plasma frequency, allowing the scalar-potential contribution associated with charge separation to be neglected.
Then, the perturbed magnetic field, electric field, and current density can be expressed in terms of the perturbed vector potential $\boldsymbol{A}_{1}$ as $\boldsymbol{B}_{1}=\nabla\times\boldsymbol{A}_{1}$, $\boldsymbol{E}_{1}=-\partial_{t}\boldsymbol{A}_{1}/c$, $\nabla\times\boldsymbol{B}_{1}=4\pi\boldsymbol{J}_{1}/c$. These expressions are valid in both the ideal MHD and non-ideal MHD regimes.

The expression for the perturbed vector potential in the ideal-MHD region was derived by White \cite{White86},
\begin{equation}
    \tilde{A}_{z1}(x)=\tilde{a}_{k}\left(1+\frac{\tanh\left(|x|/w\right)}{kw}\right)\exp\left(-k|x|\right) 
    \label{ideal_mhd_A}
\end{equation}
originally within the framework of non-relativistic resistive MHD. 
Here we show that the same equation can also be obtained for the present relativistically hot collisionless plasma. To demonstrate this, we start from the Vlasov equation and linearize it with respect to first-order perturbations. The linearized equation is written as
\begin{equation}
    \begin{split}
        \left(\frac{\partial}{\partial t}+\boldsymbol{v}\cdot\frac{\partial}{\partial \boldsymbol{x}}+\frac{q_{s}}{c}\left(\boldsymbol{v}\times\boldsymbol{B}_{0}\right)\cdot\frac{\partial}{\partial\boldsymbol{p}}\right)f_{s1}&\\
        =-q_{s}\left(\boldsymbol{E}_{1}+\frac{\boldsymbol{v}}{c}\times\boldsymbol{B}_{1}\right)&\cdot\frac{\partial f_{s0}}{\partial \boldsymbol{p}}
    \end{split}
    \label{blasov_1}
\end{equation}
The left-hand side operator corresponds to the Liouville operator with the equilibrium phase-space flow. Therefore, by introducing the characteristic of the equilibrium orbit, the linearized Vlasov equation may be expressed as a first-order ordinary differential equation along the unperturbed particle trajectory. Then, we rewrite the left-hand side of Eq.\,\eqref{blasov_1} as 
\begin{equation}
    \frac{D f_{s1}}{Dt},
\end{equation}
where $D/Dt$ denotes the total time-derivative along the unperturbed particle orbit under the equilibrium field in the phase-space.
Using the relations between the perturbed vector potential and the perturbed electromagnetic fields, the right-hand side of Eq.~\eqref{blasov_1} becomes
\begin{equation}
    \begin{split}
            &-q_{s}\left(\boldsymbol{E}_{1}+\frac{\boldsymbol{v}}{c}\times\boldsymbol{B}_{1}\right)\cdot\frac{\partial f_{s0}}{\partial \boldsymbol{p}}=
            -q_{s}\left[\frac{i\omega}{c}A_{z1}\mathbf{e}_{z}+\exp\left(iky-i\omega t\right)\right.\\ 
             &\left\{\left.\frac{ik\tilde{A}_{z1}\left(x\right)}{c}\left(v_{z}\mathbf{e}_{y}-v_{y}\mathbf{e}_{z}\right)-\frac{1}{c}\frac{\partial\tilde{A}_{z1}\left(x\right)}{\partial x}\left(v_{x}\mathbf{e}_{z}- v_{z}\mathbf{e}_{x}\right)\right\}\right]\\
             &\hspace{5cm}
             \times\left(-\frac{c^{2}} {T_{s}E_{s}}\boldsymbol{p}+\frac{c\beta_{s}}{T_{s}}\mathbf{e}_{z}\right)f_{s0}\\
             &=q_{s}f_{s0}\left(\frac{i\omega}{cT_{s}}v_{z}A_{z1}-\frac{\beta_{s}}{T_{s}}i\omega A_{z1}+\frac{\beta_{s}}{T_{s}}ikv_{y}A_{z1}\right.\\ 
             &\hspace{4cm} 
             +\left.\frac{\beta_{s}}{T_{s}}v_{x}\frac{\partial \tilde{A}_{{z1}}\left(x\right)}{\partial x}\exp\left(iky-i\omega t\right) \right)\\
             &=\frac{q_{s}f_{s0}}{cT_{s}}\left(c\beta_{s}\frac{dA_{z1}}{dt}+i\omega\boldsymbol{v}\cdot\boldsymbol{A}_{1}\right).
    \end{split}
\end{equation}
In deriving the above expression, we have used $p_{z}E_{s}=v_{z}/c^{2}$, $p_{y}v_{z}=p_{z}v_{y}$, and $p_{z}v_{x}=p_{x}v_{z}$. Combining the expression for the left- and right-hand sides, we obtain
\begin{equation}
    \frac{Df_{s1}}{Dt}=\frac{q_{s}f_{s0}}{cT_{s}}\left(c\beta_{s}\frac{dA_{z1}}{dt}+i\omega\boldsymbol{v}\cdot\boldsymbol{A}_{1}\right)
    \label{blasov_2}
\end{equation}
Integrating Eq.~\eqref{blasov_2} along the unperturbed particle orbit from $t'=-\infty$ to $t'=t$, we obtain
\begin{equation}
    f_{s1}=\frac{q_{s}f_{s0}}{cT_{s}}\left(c\beta_{s}A_{z1}+i\omega\int_{-\infty}^{t}\boldsymbol{v}\cdot\boldsymbol{A}_{1}dt^{\prime}\right)
    \label{blasov_3}
\end{equation}
As discussed in Ref.~\cite{Coppi66,Hoshino20,Hoshino21}, the first term on the right-hand side of Eq.~\eqref{blasov_3} corresponds to the ideal-MHD response, whereas the second term represents the non-ideal kinetic contribution. Retaining only the ideal-MHD term, the perturbed current density becomes 
\begin{equation}
    \boldsymbol{J}_{1}^{\rm ideal}=\int d^{3}p\sum_{s}q_{s}v_{z}\frac{q_{s}f_{s0}}{cT_{s}}c\beta_{s}A_{z1}\mathbf{e}_{z}.
    \label{ideal_current}
\end{equation}
Substituting this current into Amp{\`e}re's law without the displacement-current term,
$\nabla\times\left(\nabla\times\boldsymbol{A}_{1}\right)=4\pi\boldsymbol{J}_{1}^{\rm ideal}/c$
and considering the form $\boldsymbol{A}_{1}=\tilde{A}_{z1}(x)\exp\left(iky-i\omega t\right)\mathbf{e}_{z}$,
with $\tilde{A}_{z1}(x)$ given by Eq.~\eqref{ideal_mhd_A}, 
the left-hand side becomes
\begin{equation}
    \nabla\times\left(\nabla\times\boldsymbol{A}_{1}\right)=-\left\{\frac{d^{2}\tilde{A}_{z1}(x)}{dx^{2}}-k^{2}\tilde{A}_{z1}(x)\right\}\exp\left(iky-i\omega t\right)\mathbf{e}_{z}
    \label{ampere_left1}
\end{equation}
and the right-hand side does
\begin{equation}
    \begin{split}
    &\frac{4\pi\boldsymbol{J}_{1}^{\rm ideal}}{c}=\frac{4\pi}{c}\int d^{3}p\sum_{s}q_{s}v_{z}\frac{q_{s}f_{s0}}{cT_{s}}c\beta_{s}A_{z1}\mathbf{e}_{z}\\
    &=4\pi e^{2}cA_{z1}\left(\int\limits_{s=e} d^{3}p\frac{p_{z}f_{e0}}{T_{e}E_{e}}\beta_{e}+\int\limits_{s=p} d^{3}p\frac{p_{z}f_{p0}}{T_{p}E_{p}}\beta_{p}\right)\mathbf{e}_{z}.
    \end{split}
    \label{ampere_right1}
\end{equation}
The integrals can be evaluated as 
\begin{equation}
    \begin{split}
    &\int\limits_{s}d^{3}p\frac{\beta_{s}p_{z}}{T_{s}E_{s}}\frac{\overline{n}_{s0}}{4\pi m_{s}^{2}c\Theta_{s}K_{2}\left(m_{s}c^{2}/\Theta_{s}\right) }\\
    &{\hspace{3cm}}\times\exp\left(-\frac{ E_{s}-c\beta_{s}p_{z}-q_{s}\beta_{s}A_{z0}(x) }{T_{s}}\right)\\
    &=\int\limits_{s}d^{3}p\frac{\beta_{s}p_{z}}{T_{s}E_{s}}\frac{\overline{n}_{s0}}{4\pi m_{s}^{2}c\Theta_{s}K_{2}\left(m_{s}c^{2}/\Theta_{s}\right) }\\
    &{\hspace{3cm}}\times\exp\left(-\frac{ \overline{E}_{s} }{\Theta_{s}}\right)\exp\left(\frac{ q_{s}\beta_{s}A_{z0}(x) }{T_{s}}\right)\\
    &=\frac{\beta_{s}}{T_{s}}\exp\left(\frac{ q_{s}\beta_{s}A_{z0}(x) }{T_{s}}\right)\int\limits_{s}d^{3}p\frac{p_{z}}{E_{s}}\overline{f}_{s,M}\left(\overline{\boldsymbol{p}}\right)\\
    &=\frac{\beta_{s}}{T_{s}}\exp\left(\frac{ q_{s}\beta_{s}A_{z0}(x) }{T_{s}}\right)\int\limits_{s}d^{3}\overline{p}\frac{\Gamma_{s}\left(\overline{p}_{z}+\beta_{s}\overline{E}_{s}/c\right)}{\overline{E}_{s}}\overline{f}_{s,M}\left(\overline{\boldsymbol{p}}\right)\\
    &=\frac{\beta_{s}^{2}n_{s0}}{cT_{s}}\exp\left(\frac{ q_{s}\beta_{s}A_{z0}(x) }{T_{s}}\right),
    \end{split}
    \label{int_ideal_j}
\end{equation}
where variables with an overbar denote quantities measured in the bulk rest frame. $\overline{f}_{s,M}$ and $\overline{E}_{s}=\Gamma_{\beta}\left(E_{s}-c\beta_{s}p_{z}\right)$ represent the relativistic Maxwellian distribution and relativistic particle energy for the $s$-species in that frame, respectively. In the above derivation, we perform a change of variables based on the Lorentz transformation and the identity $d^{3}p/E_{s}=d^{3}\overline{p}/\overline{E}_{s}$\cite{landau1975classical} has been employed. Assuming $n_{e0}=n_{p0}=n_{0}$, the Harris equilibrium satisfies $\beta_{p}/T_{p}=-\beta_{e}/T_{e}$\cite{Hoh66}. Assuming the pair plasma configuration adopted in the simulation setup of this paper, in which the electrons and positrons have equal temperatures, $T_{p}=T_{e}=T$, and equal-magnitude drift velocities in opposite directions, $\beta_{p}=-\beta_{e}=\beta$, the equilibrium vector potential takes the form
\begin{equation}
    A_{z0}(x)=-\frac{2T}{e\beta}\ln\,\cosh\left(\beta\sqrt{\frac{4\pi n_{0}e^{2}}{T}}x\right).
    \label{equiv_Az}
\end{equation}
The sheet thickness $w$ can be written as $w=\sqrt{T/4\pi n_{0}e^{2}}/\beta$.
Substituting Eqs.~\eqref{int_ideal_j} and \eqref{equiv_Az} into to the righ-hand side of Eq.~\eqref{ampere_right1}, and using the Harris equilibrium conditions, $B_{0}^{2}/8\pi=2n_{0}T$ and $2T/w=e\beta B_{0}$, we obtain 
\begin{equation}
    \begin{split}
    \frac{4\pi\boldsymbol{J}_{1}^{\rm ideal}}{c}&=\frac{8\pi n_{0}e^{2}\beta^{2}}{T}\cosh^{-2}\left(\frac{x}{w}\right)A_{z1}\mathbf{e}_{z}\\
    &=\frac{2}{w^{2}}\cosh^{-2}\left(\frac{x}{w}\right)A_{z1}\mathbf{e}_{z}
    \end{split}
    \label{ampere_right2}
\end{equation}
Combining this result with Eq.~\eqref{ampere_left1}, we arrive at
\begin{equation}
    \frac{d^{2}\tilde{A}_{z1}(x)}{d x^{2}}-k^{2}\tilde{A}_{z1}(x)=-\frac{2}{w^{2}}\cosh^{-2}\left(\frac{x}{w}\right)\tilde{A}_{z1}(x).
    \label{blasov_4}
\end{equation}
This equation is identical to the Eq.~(23) derived by White in Ref.~\cite{White86}. Therefore, the eigenfunction in the ideal-MHD region is also given by the same solution, namely Eq.~\eqref{ideal_mhd_A}, even for the present relativistically hot collisionless plasma.

Using Eqs.~\eqref{ideal_mhd_A}, \eqref{ampere_right2}, and $\boldsymbol{E}_{1}=-\partial_{t}\boldsymbol{A}_{1}/c$, the integrands on the left-hand side of Eq.\,\eqref{energy_eq} can be evaluated as
\begin{equation}
\begin{split}
&\frac{\partial}{\partial t}\exp(2\gamma t)|\tilde{\boldsymbol{B}_{1}}(x)|^{2}=2\gamma \exp\left(2\gamma t\right)\left(ik\tilde{A}_{z1}(x)\mathbf{e}_{x}-\frac{d\tilde{A}_{z1}(x)}{dx}\mathbf{e}_{y}\right)\\
&\hspace{4cm}\times\left(-ik\tilde{A}_{z1}^{*}(x)\mathbf{e}_{x}-\frac{d\tilde{A}_{z1}^{*}(x)}{dx}\mathbf{e}_{y}\right)\\
&=\left(k^{2}|\tilde{A}_{z1}(x)|^{2}+\left|\frac{d\tilde{A}_{z1}(x)}{dx}\right|^{2}\right)\exp\left(2\gamma t\right),
\end{split}
\end{equation}
and 
\begin{equation}
    \begin{split}
    &{\rm Re}\left[\tilde{E}_{z1}^{*}(x)\tilde{J}_{z1}^{\rm ideal}(x)\right]\exp\left(2\gamma t\right)\\
    &={\rm Re}\left[-\frac{\gamma}{c}\tilde{A}_{z1}^{*}\left(x\right)\times\frac{c}{2\pi w^{2}\cosh^{2}(x/w)}\tilde{A}_{z1}\left(x\right)\right]\\
    &=-\frac{\gamma}{2\pi w^{2}\cosh^{2}(x/w)}\left|\tilde{A}_{z1}(x)\right|^{2}\exp\left(2\gamma t\right),
    \end{split}
\end{equation}
respectively.
Then, the left-hand side of Eq.\,\eqref{energy_eq} can be written as\cite{Hoshino20,Hoshino21}
\begin{equation}
    \begin{aligned}
   &\int\limits_{x=-\infty}^{x=\infty}\left\{\frac{\partial}{\partial t}\left(\frac{|\tilde{\boldsymbol{B}}_{1}(x)|^{2}}{8\pi}{\rm e}^{2\gamma t}\right)+{\rm Re}\left[\tilde{E}_{z1}^{*}(x)\tilde{J}_{z1}^{\rm ideal}(x)\right]{\rm e}^{2\gamma t}\right\}dx\\
   &=\frac{\gamma\exp\left(2\gamma t\right)}{4\pi}\int\limits_{x=-\infty}^{x=\infty}\left\{k^{2}|\tilde{A}_{z1}(x)|^{2}+\left|\frac{d\tilde{A}_{z1}(x)}{dx}\right|^{2}\right.\\
   &\hspace{5cm}
   -\left.\frac{\left|\tilde{A}_{z1}(x)\right|^{2}}{w^{2}\cosh^{2}(x/w)}\right\}dx\\
   &=\frac{\gamma\exp\left(2\gamma t\right)}{4\pi}\int\limits_{x=-\infty}^{x=\infty}\left\{\left(\frac{d\tilde{A}_{z1}}{dx}\right)^{*}\left(\frac{d\tilde{A}_{z1}}{dx}\right)+\tilde{A}_{z1}^{*}(x)\frac{d^{2}\tilde{A}_{z1}\left(x\right)}{d x^{2}}\right\}dx\\
   &=\frac{\gamma\exp\left(2\gamma t\right)}{4\pi}\int\limits_{x=-\infty}^{x=\infty}\frac{d}{d x}\left(\tilde{A}_{z1}^{*}(x)\frac{d \tilde{A}_{z1}(x)}{dx}\right)dx\\
       &=\frac{\gamma}{2\pi}\frac{\left(k^{2}w^{2}-1\right)}{kw^{2}}\left| \tilde{a}_{k}\right|^{2}{\rm exp(2\gamma t)}.
    \end{aligned}
    \label{ideal_mhd_energy}
\end{equation}
Here, we have used that $\tilde{A}_{z1}^{*}(x)d^{2}\tilde{A}_{z1}(x)/dx^{2}=k^{2}|\tilde{A}_{z1}(x)|^{2}-2|\tilde{A}_{z1}(x)|^{2}/w^{2}\cosh^{2}(x/w).$ The factor $\gamma$ originates from the temporal variation of the magnetic energy density and from the relation $\boldsymbol{E}_{1}=-\partial\boldsymbol{A}_{1}/c\partial t$.

The perturbed current density ${J}_{1}^{\rm non-ideal}$ in the right-hand side of Eq.~\eqref{energy_eq} can be represented as
\begin{equation}
    {J}_{1}^{\rm non-ideal}=\int d^{3}p\sum_{s}q_{s}v_{z}\frac{q_{s}f_{s0}}{cT_{s}}\left(i\omega\int_{-\infty}^{t}\boldsymbol{v}\cdot\boldsymbol{A}_{1}dt^{\prime}\right),
    \label{non_ideal_current}
\end{equation}
which is obtained from the second term on the right-hand side of Eq.~\eqref{blasov_3}.
With the particle trajectory inside the non-ideal MHD region approximated by the straight-line orbit $x={\rm const.}$ and $y=v_{y}t$, the integral in Eq.~\eqref{non_ideal_current} can be evaluated as 
\begin{equation}
    \begin{split}
        &i\omega\int\limits_{-\infty}^{t}\boldsymbol{v}\cdot\boldsymbol{A}_{1}dt^{\prime}=i\omega\int\limits_{-\infty}^{t}v_{z}\tilde{A}_{z1}\left(x^{\prime}\right)\exp\left(iky^{\prime}-i\omega t^{\prime}\right)dt^{\prime}\\
        &=i\omega\int\limits_{-\infty}^{t}v_{z}\tilde{A}_{z1}\left(x^{\prime}\right)\exp\left[ik\left\{y-v_{y}\left(t-t^{\prime}\right)\right\}-i\omega\left(t^{\prime}-t\right)-i\omega t\right]dt^{\prime}\\
        &=i\omega v_{z}\tilde{A}_{z1}\left(x\right)\exp\left(iky-i\omega t\right)\int\limits_{-\infty}^{0}\exp \left\{i\left(kv_{y}-\omega\right)\tau\right\}d\tau\\
        &=i\omega v_{z}\tilde{A}_{z1}\left(x\right)\exp\left(iky-i\omega t\right){\lim_{\epsilon \to +0}}\int\limits_{-\infty}^{0}\exp \left\{\epsilon\tau+i\left(kv_{y}-\omega\right)\tau\right\}d\tau\\
        &=-\frac{\omega}{\omega-kv_{y}+i\epsilon_{0}}v_{z}A_{z1}H\left(x\right),
    \end{split}
\end{equation}
where variables with a prime represent quantities at time $t^{\prime}$ and $\tau$ is defined as $\tau\equiv t^{\prime}-t$. $\epsilon_{0}$ indicates the limit in which $\epsilon$ approaches zero from the positive side. Since only particles executing meandering orbits contribute appreciably to the non-adiabatic response, we approximate the response as being confined to the meandering region. This spatial localization is represented by a Heaviside function $H(x)$,
\begin{equation}
   H(x)=\begin{cases}
        1 & (|x|\le r_{m})\\
        0 & (|x|> r_{m})
    \end{cases}.
\end{equation}
To simplify the evaluation of Eq.~\eqref{non_ideal_current}, we assume that the equilibrium vector potential is approximately zero in the region ($|x|\le r_{m}$), since this region is localized around the origin, where Eq.~\eqref{equiv_Az} indicate that the equilibrium vector potential is nearly zero. Under this approximation, Eq.~\eqref{non_ideal_current} reduces to the following expression for $|x|< r_{m}$,
\begin{equation}
\begin{split}
    &{J}_{1}^{\rm non-ideal}=-\sum_{s} \int d^{3}p\frac{q_{s}^{2}\omega A_{z1}}{cT_{s}}\frac{\overline{n}_{s0}}{4\pi m_{s}^{2}c\Theta K_{2}\left(m_{s}c^{2}/\Theta\right)}\\
    &\hspace{2cm}\times\exp\left(-\frac{\Gamma_{s}\left(E_{s}-c\beta_{s}p_{z}\right)}{\Theta}\right)\frac{v_{z}^{2}}{\omega - kv_{y}+i\epsilon_{0}}\\
    &=-\sum_{s} \int \frac{d^{3}p}{E_{s}}E_{s}\frac{q_{s}^{2}\omega A_{z1}}{cT_{s}}\frac{\overline{n}_{s0}}{4\pi m_{s}^{2}c\Theta K_{2}\left(m_{s}c^{2}/\Theta\right)}\\
    &\hspace{4cm}\times\exp\left(-\frac{\overline{E}_{s}}{\Theta}\right)\frac{v_{z}^{2}}{\omega - kv_{y}+i\epsilon_{0}}\\
    &=-\sum_{s} \int \frac{d^{3}\overline{p}}{\overline{E_{s}}}\Gamma_{s}\left(\overline{E_{s}}+c\beta_{s}\overline{p}_{z}\right)\frac{q_{s}^{2}\omega A_{z1}}{cT_{s}}\frac{\overline{n}_{s0}}{4\pi m_{s}^{2}c\Theta K_{2}\left(m_{s}c^{2}/\Theta\right)}\\
    &\hspace{4cm}\times\exp\left(-\frac{\overline{E}_{s}}{\Theta}\right)\frac{v_{z}^{2}}{\omega - kv_{y}+i\epsilon_{0}}\\
   & =-\sum_{s} \int d^{3}\overline{p}\frac{q_{s}^{2}\omega A_{z1}}{cT_{s}}\frac{\Gamma_{s}\overline{n}_{s0}}{4\pi m_{s}^{2}c\Theta K_{2}\left(m_{s}c^{2}/\Theta\right)}\\
    &\hspace{4cm}\times\exp\left(-\frac{\overline{E}_{s}}{\Theta}\right)\frac{v_{z}^{2}}{\omega - kv_{y}+i\epsilon_{0}}.
    \end{split}
    \label{non_ideal_current2}
\end{equation}
In deriving Eq.~\eqref{non_ideal_current2}, we perform a change of variables based on the Lorentz transformation and make use of the relation $d^{3}p/E_{s}=d^{3}\overline{p}/\overline{E}_{s}$. Since the plasma temperature is assumed to be highly relativistic, the particle velocity along the $z$-direction satisfies $v_{z}^{2}\sim c^{2}$.
Consequently, Eq.~\eqref{non_ideal_current2} can be further simplified as
\begin{equation}
\begin{split}
   J_{1}^{\rm non-ideal} =-\sum_{s} \int d^{3}\overline{p}\frac{q_{s}^{2}\omega A_{z1}}{cT_{s}}\frac{\Gamma\overline{n}_{s0}}{4\pi m_{s}^{2}c\Theta K_{2}\left(m_{s}c^{2}/\Theta\right)}\\
    \times\exp\left(-\frac{\overline{E}_{s}}{\Theta}\right)\frac{c^{2}}{\omega - kv_{y}+i\epsilon_{0}}.
    \end{split}
\end{equation}
Applying the Plemelj formula and relating only the resonant contribution, we obtain\cite{Hoshino20}
\begin{equation}
\begin{split}
   &J_{1}^{\rm non-ideal} =\sum_{s} \frac{q_{s}^{2}\omega A_{z1}}{cT_{s}}\frac{\Gamma^{2}_{s}\overline{n}_{s0}}{4\pi m_{s}^{2}c\Theta K_{2}\left(m_{s}c^{2}/\Theta\right)}\\
&\times2i\frac{\pi^{2}}{k}m_{s}^{3}c^{4}\exp\left(-\frac{m_{s}c^{2}}{\Theta}\right)\frac{\Theta}{m_{s}c^{2}}\left\{1+\frac{2\Theta}{m_{s}c^{2}}\left(1+\frac{\Theta}{m_{s}c^{2}}\right)\right\}.
\end{split}
\label{non_ideal_current3}
\end{equation}
Finally, assuming a sufficiently relativistic plasma temperature and approximating the modified Bessel function of the second kind\cite{Zelenyi79} by $K_{2}\left(m_{s}c^{2}/\Theta_{s}\right)\simeq 2\left(\Theta_{s}/m_{s}c^{2}\right)^{2}$, Eq.~\eqref{non_ideal_current3} becomes
\begin{equation}
    \begin{split}
    J_{1}^{\rm non-ideal}&\simeq\sum_{s}\frac{q_{s}^{2}\omega A_{z1}}{cT_{s}}\frac{\Gamma_{s}^{2}\overline{n}_{s0}}{4\pi m_{s}^{2}c}\times 2i\frac{\pi^{2}}{k}m_{s}^{2}c^{2}\\
    &=-\frac{\gamma\pi e^{2}n_{0}\Gamma_{\beta}}{kT}A_{z1}.
    \end{split}
\end{equation}

An analytical solution for the perturbed vector potential in the non-ideal MHD region cannot be obtained. 
Previous studies therefore assumed that the spatial profile of the perturbation is uniform within the non-ideal MHD region; this is known as the constant-A approximation. 
Under the constant-A approximation, the right-hand side of Eq.\,\eqref{energy_eq} is evaluated as\cite{Hoshino20}
\begin{equation}
    \begin{split}
       -\int\limits_{x=-r_{m}}^{x=r_{m}} &{\rm Re}\left[\tilde{E}_{z1}^{*}(x)\tilde{J}_{z1}^{\rm non-ideal}(x)\right]{\rm e}^{2\gamma t}dx\\
       &\simeq-\pi \frac{m_{0}c^{2}}{\Theta}\frac{\overline{n}_{0}e^{2}}{m_{0}}\frac{\Gamma^{3}_{\beta}}{kc}\frac{\gamma^{2}{\rm exp(2\gamma t)}}{c^{2}}\int\limits_{x=-r_{m}}^{x=r_{m}}\tilde{A}_{z1}^{*}(x)\tilde{A}_{z1}(x)dx\\
       &\simeq-\pi \frac{m_{0}c^{2}}{\Theta}\frac{\overline{n}_{0}e^{2}}{m_{0}}\frac{\Gamma^{3}_{\beta}}{kc}\frac{\gamma^{2}}{c^{2}}\left|\tilde{a}_{k}\right|^{2}{\rm exp(2\gamma t)}w\sqrt{\frac{\beta}{2}}.
    \end{split}
    \label{non_ideal_mhd_energy_Constant_A}
\end{equation}
Previous studies\cite{Zenitani01,Bessho07} have shown that particle acceleration by the electric field in the non-ideal MHD region occurs precominantly in the vicinity of the X-point. Because the magnetic field is relatively weak there, particles become only weakly magnetized and can therefore be efficiently accelerated by the electric field. Although  an electric field is also induced around the O-point, the stronger magnetic field in this region keeps particles more tightly magnetized, so that the contribution of electric field acceleration is expected to be much smaller than around X-point. In deriving the energy equation, Eq.~\eqref{eq_energy_conservation}, the integration over the $y$ direction was performed over one perturbation wavelength. As a simplified estimate of the contribution from the non-ideal MHD region, however, we assume that only the half-wavelength region surrounding the X-point contributes significantly to particle acceleration. Accordingly, the right-hand side of Eq.~\eqref{non_ideal_mhd_energy_Constant_A} is multiplies by a factor of $1/2$.

Combining Eqs.\,\eqref{ideal_mhd_energy} and \eqref{non_ideal_mhd_energy_Constant_A}, we obtain the following expression for the growth rate:
\begin{equation}
    \gamma(k) =\frac{2\sqrt{2}}{\pi}\frac{1}{\tau_{c}}\left(1-k^{2}w^{2}\right)\frac{\beta^{3/2}}{\Gamma_{\beta}}.
    \label{growth_Constant_A}
\end{equation}
In the previous study\cite{Hoshino20}, an additional factor of $kw$ was introduced on the right-hand side so that the growth rate vanishes at $x=0$. With this factor included, the expression reduces to Eq.\,\eqref{eq_growth_Hoshino}, consistent with the result obtained in the previous study.

In principle, however, the perturbation in the non-ideal MHD region should be smoothly matched that in the ideal MHD region. 
For modes with wavenumber $k$ close to unity, the spatial gradient of the perturbed vector potential near the boundary of the non-ideal MHD region is relatively small, and the constant-A approximation is therefore well justified.

In contrast, for modes with wavenumber $kw$ close to zero, the spatial gradient of the perturbed vector potential becomes steep, and the constant-A approximation becomes worse. 
Instead of assuming a spatially uniform perturbation within the non-ideal MHD region, the perturbation solution derived in the ideal-MHD region is linearly extrapolated into the non-ideal region and used as an effective approximation there. This assumption corresponds to the extrapolated-A approximation introduced in the main text.

The integration 
\begin{equation*}
    \int\limits_{x=-r_{m}}^{x=r_{m}}\tilde{A}_{z1}^{*}(x)\tilde{A}_{z1}(x)dx,
\end{equation*}
as shown in the derivation of Eq.~\eqref{non_ideal_mhd_energy_Constant_A}, that arises in the evaluation of the right-hand side of Eq.~\eqref{energy_eq} can be evaluated only over the half-domain ($x\ge0$) by virtue of the symmetry of $|\tilde{A}_{z1}(x)|$, thereby simplifying the handling of the integration.
Because the non-ideal region is confined to $x/w \ll 1$, the solution there is approximated by a linear extrapolation of the ideal-MHD solution. Accordingly, expanding Eq.\,\eqref{ideal_mhd_A} to first order in $x$ yields $\tilde{A}_{z1}\left(x\right) \sim \tilde{a}_{k}\left(1+x/kw^{2}\right)$.
With the extrapolated-A approximation, the spatial integral of the perturbed vector potential appearing in the first line on the right-hand side of Eq.~\eqref{non_ideal_mhd_energy_Constant_A} can be rewritten as
\begin{equation}
    \begin{split}
     2\int\limits_{x=0}^{x=r_{m}}\left|\tilde{A}_{z1}(x)\right|^{2}dx&\simeq2\int\limits_{x=0}^{x=r_{m}}|a_{k}|^{2}\left(1+\frac{x}{kw^{2}}\right)^{2}dx \\
     &\simeq 2r_{m}|a_{k}|^{2} \frac{1}{kw}\left(kw+\frac{r_{m}}{w}\right).
    \end{split}
    \label{eq_A_2}
\end{equation}
Compared with the integral value $2r_{m}|a_{k}|^{2}$ obtained under the constant-A approximation, the present result differs only by an additional factor $\left(kw+r_{m}/w\right)/kw$. Consequently, multiplying Eq.~\eqref{non_ideal_mhd_energy_Constant_A} by this factor and equating the result to the right-hand side of Eq.~\eqref{ideal_mhd_energy} yields the growth rate given by Eq.~\eqref{eq_growth_New}.

Importantly, the factor $kw$, which was phenomenologically multiplied by $\left(1-k^{2}w^{2}\right)$ in previous study \cite{Hoshino20}  as correction to transform Eq.\,\eqref{growth_Constant_A} into Eq.\,\eqref{eq_growth_Hoshino} and ensure that the growth rate vanishes at $k=0$, is naturally reproduced in the present formulation.
The meandering orbit radius of relativistic particles can be expressed in the same functional form as in the non-relativistic case\cite{Sonnerup71}. 
However, the particle mass is effectively multiplied by the Lorentz factor, as discussed in the Appendix \ref{App_A}. By inserting the relativistic meandering radius Eq.\,\eqref{eq_rela_meandering} into $r_{m}$ in Eq.\,\eqref{eq_A_2}, the growth rate can be evaluated for the relativistic regime, and we obtain Eq.~\eqref{eq_growth_New}.

\nocite{*}
\bibliography{aipsamp}

\end{document}